\documentclass[12pt,preprint]{aastex}

\usepackage{graphicx}
\usepackage{amssymb}
\begin{document}

\title{H$_2$ Emission Nebulosity Associated with KH 15D\altaffilmark{1}}

\author{
A. T. Tokunaga,\altaffilmark{2} 
S. Dahm,\altaffilmark{2}
W. G{\"a}ssler,\altaffilmark{3}
Yutaka Hayano,\altaffilmark{4}
Masahiko Hayashi,\altaffilmark{5}
Masanori Iye,\altaffilmark{4}
Tomio Kanzawa,\altaffilmark{5}
Naoto Kobayashi,\altaffilmark{6}
Yukiko Kamata,\altaffilmark{4}
Yosuke Minowa,\altaffilmark{6}
Ko Nedachi,\altaffilmark{5}
Shin Oya,\altaffilmark{5}
Tae-Soo Pyo,\altaffilmark{5}
D. Saint-Jacques,\altaffilmark{7}
Hiroshi Terada,\altaffilmark{5} 
Hideki Takami,\altaffilmark{5} and 
Naruhisa Takato\altaffilmark{5}  }

\altaffiltext{1} {Based on data collected at Subaru Telescope, which 
   is operated by the National Astronomical Observatory of Japan.}

\altaffiltext{2}{Institute for Astronomy, University of Hawaii, 2680 Woodlawn 
   Drive, Honolulu, HI 96822;
   tokunaga@ifa.hawaii.edu, dahm@ifa.hawaii.edu.}

\altaffiltext{3}{Max-Planck-Institut f\"ur Astronomie, K\"onigstuhl 17,
Heidelberg D-69117, Germany; gaessler@mpia-hd.mpg.de}

\altaffiltext{4}{National Astronomical Observatory of Japan, Mitaka, 
   Tokyo 181-8588, Japan; y.hayano@nao.ac.jp, iye@optik.mtk.nao.ac.jp,
   kamata@merope.mtk.nao.ac.jp}

\altaffiltext{5}{Subaru Telescope, 650 North A`ohoku Place, Hilo, HI  96720;
Firstname.Lastname@SubaruTelescope.org.}

\altaffiltext{6}{Institute of Astronomy, Graduate School of Science, University
of Tokyo, Mitaka-shi, Tokyo 181-0015, Japan; naoto@ioa.s.u-tokyo.ac.jp.}

\altaffiltext{7}{Groupe d'astrophysique, Universit{\'e} de Montr{\'e}al,
2900 Boul. {\'E}douard-Montpetit, Montr{\'e}al, Qu{\'e}bec, H3T 1J4 Canada;
dastj14@agora.ulaval.ca.}

\begin{abstract}

An H$_2$ emission filament is found in close proximity to the
unique object KH~15D using the adaptive optics system of the Subaru Telescope.  
The morphology of the filament, the presence of spectroscopic outflow signatures
observed by Hamilton et al., and the detection of extended H$_2$ 
emission from KH~15D by Deming, Charbonneau, \& Harrington suggest that 
this filament arises from shocked H$_2$ in an outflow.   
The filament extends about 15$''$ to the north of KH 15D.

\end{abstract}


\section{Introduction}

KH 15D is an eclipsing  weak-lined T Tauri star with remarkable properties:  
(1) the eclipse has a  48 or 96 day period with a depth of 3.5 mag; 
(2) the eclipses have been lengthening (16 days in 1995 to about 24
days currently), thus requiring obscuring matter to extend
at least one-third of the circumference of the orbit (Hamilton et al. 2001;
Herbst et al. 2002; Winn et al. 2003); 
(3) polarization measurements suggest that only
scattered light is observed during the eclipse (Agol et al. 2003); 
(4) broad emission lines during the eclipse show that it is driving an 
outflow  (Hamilton et al. 2003); and 
(5) extended H$_2$ emission is observed during its eclipse 
(Deming, Charbonneau, \& Harrington 2004).
Herbst et al. (2002) find that the light curve indicates that the edge of the
obscuring matter could be modeled as a knife edge.  
The photometric colors do not change as the eclipse progresses, and this
indicates that the obscuring matter is opaque or the particles are large.
The latter is supported by the polarization measurements.

The spectral type of KH 15D is K6V (Algol et al. 2003) or K7V (Hamilton
et al. 2001).  
The observations to date suggest that K star is the primary object 
in the KH 15D system and
that the obscuring matter is extended and may be a protoplanetary system.  
Barge \& Viton (2003) suggest that the obscuring matter may be a dusty 
anticyclonic vortex in a gaseous disk surrounding the star with 
large particles, 1--10 cm in size, trapped within the vortex.
On the other hand, Agol et al. (2003) suggest that the obscuring matter is
a warped disk. 
Hamilton et al. (2003) emphasize that the geometry of the system is
unknown, and it has not been established whether the period of the eclipse is
48 days or twice this.

In this brief report we show that an H$_2$ emission filament is associated
with KH 15D, and it appears to show an outflow from KH 15D.

\section{Observations}

Broadband and narrowband infrared images of KH 15D were obtained 
with the Subaru Telescope
using the Infrared Camera and Spectrograph (IRCS; Tokunaga et al. 1998; 
Kobayashi et al. 2000)  during UT 2003 February  18--19.  
The Subaru Telescope adaptive optics system (Takami et al. 2003)
was used for these observations.
We used HD~47887 as a reference star for the adaptive optics system, 
since it was located 37$''$ south of KH 15D and within the working
field-of-view of the instrument.
The pixel scale was 58.00 $\pm$0.02 mas pixel$^{-1}$.

On February 18 $K$-band imaging showed a hint of a filamentary nebulosity 
that appeared to be associated with KH 15D (see Fig. 1a).
The exposure time was 90 s per exposure, and five dithered exposures were 
taken.  However the first exposure was noisy and not included. 
Thus the coadded image shown in Fig. 1a was obtained from four exposures
with a total integration time of 360 s.
The seeing was good on this night, and the full width at half-maximum (FWHM) of
the image of KH 15D after coadding was 0\farcs28.  
The image was smoothed by 5 pixels (0\farcs30) to enhance the signal
to noise of the filament.
Figures 1a and b show two knots of emission.  The center of these 
knots and their surface brightness are given in Table 1.

Follow-up imaging in the narrowband H$_2$ $\nu$=1--0 filter (2.122 $\mu$m) 
and the narrowband $K$ continuum filter (2.269 $\mu$m) on February 19 showed
conclusively that the filamentary emission is due to H$_2$ emission, as shown 
in Fig. 1c and d.  
Both filters have a FWHM of 0.032 $\mu$m.
As on the previous night, five dithered images were taken
and coadded.  The total exposure time was 900 s for the images shown in
Figures 1c and d.  The seeing was worse on February 19 and the FWHM
of the H$_2$ image of KH 15D after coadding was 0\farcs54.
The images were smoothed by 6 pixels (0\farcs35) to enhance the signal to 
noise of the filament.

Additional observations in the $K$ band were made by S. Dahm 
using the QUIRC 1--2.5~$\mu$m camera on the University of Hawaii 2.2 m 
telescope.  
This camera utilizes a 1024$\times$1024 HgCdTe array with a pixel scale of
0.189 arcsec pixel$^{-1}$.
The $K$ filter used here, as well as the one used for the Subaru observations, 
is part of the Mauna Kea Observatories near-infrared filter set defined by
Simons \& Tokunaga (2002) and Tokunaga, Simons, \& Vacca (2002).
The filter used had a center wavelength of 2.20 $\mu$m and a width of 
0.34 $\mu$m.
A coadded image was obtained from a set of nine dithered images with a total
integration time of 540 s and is shown in Fig. 2.
The filament can be seen in the highly stretched image.  
There is no evidence for any other nebulosity that might be related to the
filament seen near KH 15D.

\section{Discussion}

There is no evidence from the point-spread function that KH 15D is elongated 
or has a jet structure near it.
Although the filamentary emission appears to originate from KH~15D,
there is a possibility that this is a chance superposition of a 
Herbig-Haro (HH) object with KH 15D.
There are nearby embedded sources that could produce HH objects.
Embedded submillimeter and VLA sources discovered by 
Ward-Thompson et al. (2000) and Reipurth et al. (2003) are shown in Fig. 3.
The radio source VLA 5 is a candidate for the source of filament.
In this case the filament has just by coincidence reached
the vicinity of KH 15D (in projection) with the tip of the filament
aimed at KH 15D.  

While we cannot absolutely rule out this possibility, we believe that the
filament is associated with KH 15D for the following reasons:

\begin{enumerate}

\item  Spectroscopy by Deming et al. (2004) shows that KH 15D exhibits 
strong H$_2$ emission while in eclipse.  
Furthermore, the emission is {\it extended and is centered} on KH 15D. 
Since the slit was not aligned with the filament we cannot say for certain 
that the filament is connected to KH 15D, but the fact that the
extended H$_2$ emission is centered on KH 15D is a very strong indication
that the filament is associated with KH 15D.

\item Spectroscopy by Hamilton et al. (2003) suggested that KH 15D has
bipolar outflow.  
The H$_2$ filament may be associated with the bipolar outflow detected by
these authors.  If this is a correct assumption, our observations show
the position angle in the sky of the outflow.  
From Fig. 1 we find that the position angle from KH~15D to knot B
is 8\fdg4.
Although we do not see any clear indication of the counterjet, 
there is a hint of nebulosity to the south in rough alignment with the
filament and KH 15D (see Fig. 1a; the nebulosity is indicated by
``N''). 
This requires confirmation.

\item The region around KH 15D has a number of HH objects, some of
which can be seen in Fig. 2.  However, we do not see any indication of
additional nebulosity near KH~15D, and so there is no evidence
that the filament is part of a larger shock front as the curvature in
the filament might otherwise suggest.

\end{enumerate}

We therefore consider the possibility that the filament is a result of 
outflowing gas from KH 15D.
Although weak-lined T Tauri stars are generally thought to be largely stripped
of gas and to have very little or no accretion occurring, 
KH 15D is an exception.  
Hamilton et al. (2003) found that it has characteristics of a classical 
T Tauri star including weak accretion.
In addition, Bary, Weintraub, \& Kastner (2003) find H$_2$ emission in some 
weak-lined T Tauri stars, from which they conclude that not all 
weak-lined T Tauri stars have lost all of their gas.
Hence outflowing gas and shocked H$_2$ emission in an outflow from
KH 15D, while unexpected, is not surprising in this object.

Unlike the weak-lined T Tauri stars observed by Bary, Weintraub, \&
Kastner (2003) that exhibit X-ray emission and fluorescent H$_2$ emission, 
KH 15D does not exhibit X-ray emission.
T. Simon (2003, private communication) finds that the X-ray luminosity is 
less than 5 $\times$ 10$^{29}$ ergs s$^{-1}$ at the 3 $\sigma$ level
using archived data taken with XMM-Newton on 2002 March 17. 
The H$_2$ emission associated with KH 15D is likely to be shocked H$_2$
rather than fluorescence because of the morphology of the filament
and because Deming et al. (2004) find that the H$_2$ emission
is consistent with thermal excitation.

The outflow velocity is very high; Hamilton et al. (2003) observe velocities 
as high as $\pm$200 km s$^{-1}$ in the wings of the H$\alpha$ emission
line and $-$60 to $+$50 km s$^{-1}$ in the wings of the 
[\ion{O}{1}] $\lambda$6300 emission line.
Deming et al. (2004) find that the H$_2$ emission is as high as 
$-$63  km s$^{-1}$.
Since H$_2$ dissociates in shocks at such high velocities, 
it is likely that the H$_2$ emission arises as a result of the 
outflowing gas impacting co-moving gas that is at a slightly slower velocity
(i.e., see the case discussed by Micono et al. 1998 and the
discussion by Eisl{\"o}ffel et al. 2000).

The filament extends out to about 15$''$ from KH 15D.  
Given the distance to KH 15D, 760 pc (Park et al. 2000), 
and assuming an average outflow speed of 100 km s$^{-1}$ 
(Hirth, Mundt, \& Solf 1977), 
then the time to form the filament is 540 yr.  
This is not unusual for an HH outflow 
(Eisl{\"o}ffel et al. 2000; Reipurth \& Bally 2001 ).
What is somewhat unusual is the filamentary nature of the outflow 
and  that there is no evidence for a counterjet.
H$_2$ emission in outflows are typically observed as tight knots of emission or
bow shock structures rather than in a filamentary structure. 

The picture that emerges from the images we have obtained and the
spectroscopic results of Hamilton et al. (2003) and Deming et al. (2004) 
is that KH 15D has an outflow that is marked by H$_2$ emission.  
Extinction to KH 15D is small (Hamilton et al. 2001), so the counterjet is not
obscured by high extinction as is often the case in embedded young stars.
The presence of an H$_2$ filament only to the north may be a result of
a higher gas density to the north or a case where KH 15D is just 
outside of a molecular cloud and the outflow is impacting it only toward the
north.
The curved filamentary structure could result from a cavitylike 
structure caused by the outflow with only one side exhibiting
H$_2$ emission, or it could result from a slight deflection of the
outflow by a molecular cloud.
A more extreme case of the latter is discussed by Reipurth, Raga, 
\& Heathcote (1996).

\section{Summary}

A filamentary nebula extending from KH 15D was discovered and shown
to be an H$_2$ emission nebula.
The very close proximity and orientation toward KH 15D, coupled with 
the discovery of extended H$_2$ emission surrounding KH 15D
(Deming et al. 2004), 
indicate that the nebula is likely to be associated with KH 15D.
This is consistent with the discovery of spectroscopic signatures of an outflow
in high-resolution spectra obtained by Hamilton et al. (2003).

It is apparent that KH 15D has an active accretion disk.
Coupled with the possibility that the obscuring matter causing the 
eclipse may be a forming planetary system, 
we conclude that both disk dissipation and planet formation may
be occurring in the KH 15D system.

\acknowledgments{We thank the staff and crew of the Subaru Telescope
for their invaluable assistance in obtaining the observations reported
here.  We thank B. Reipurth, D. Deming, and T. Simon for helpful discussions.
ATT was supported by NASA Cooperative Agreement no. NCC 5-538.  
This research has made use of Aladin.
}

%\clearpage

\begin{figure}
%\vspace{-2cm}
\epsscale{0.90}
\plotone{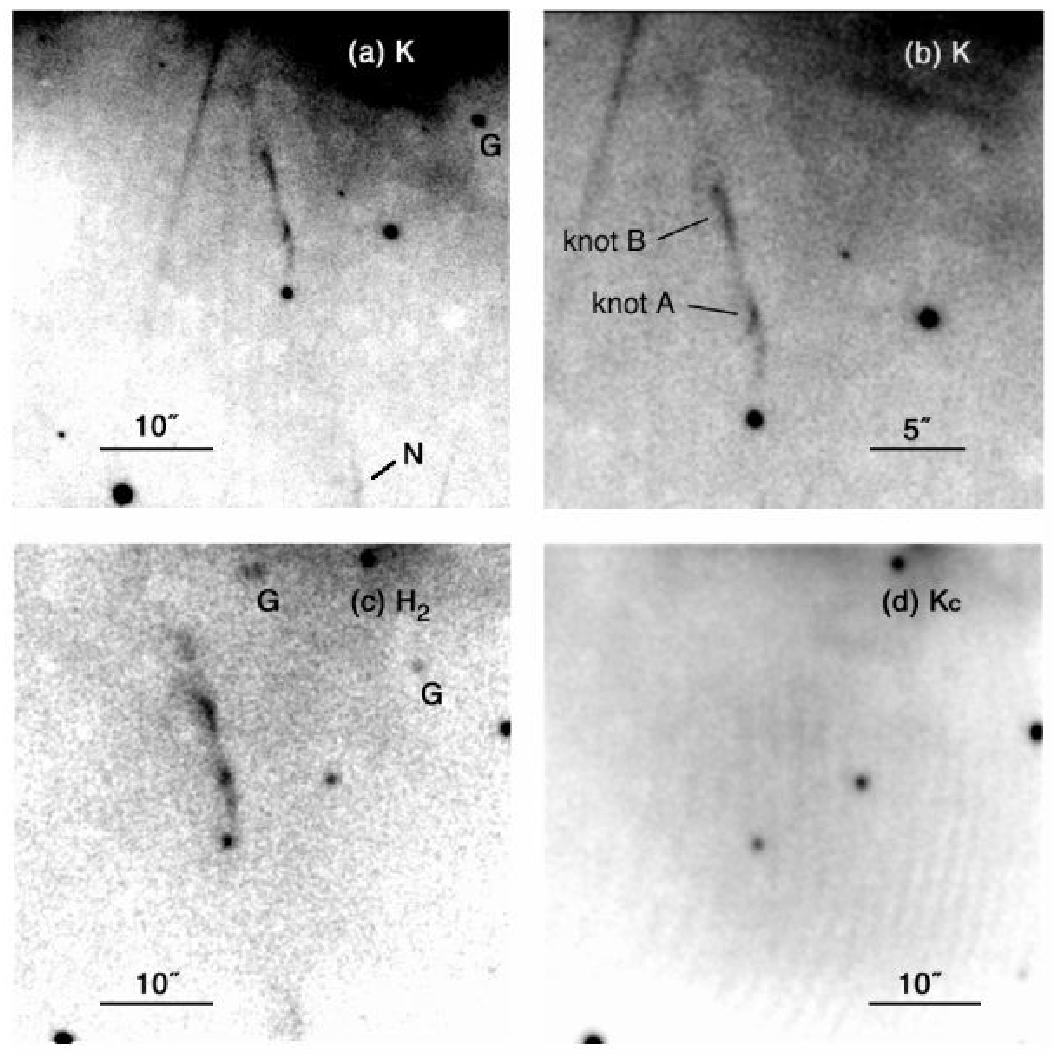}
\caption{Images of KH 15D obtained with the Subaru Telescope.  
(a) $K$-band image.  
(b) Same as (a) but expanded to show the details of the filament.
(c) Narrowband H$_2$ image.
(d) Narrowband $K$ continuum image.  
In these figures, north is up, and east is to the left.
The filament extends 6$''$ to the north at a position angle of 1$^\circ$,
then extends 7\farcs5 to the East at a position angle of 15$^\circ$.
``G" denotes ghost images that arise from the beamsplitter in IRCS.
``N" denotes faint nebulosity that may be a counterjet.
}
\end{figure}

\clearpage

\begin{figure}
%\vspace{-2cm}
%\epsscale{0.55}
\plotone{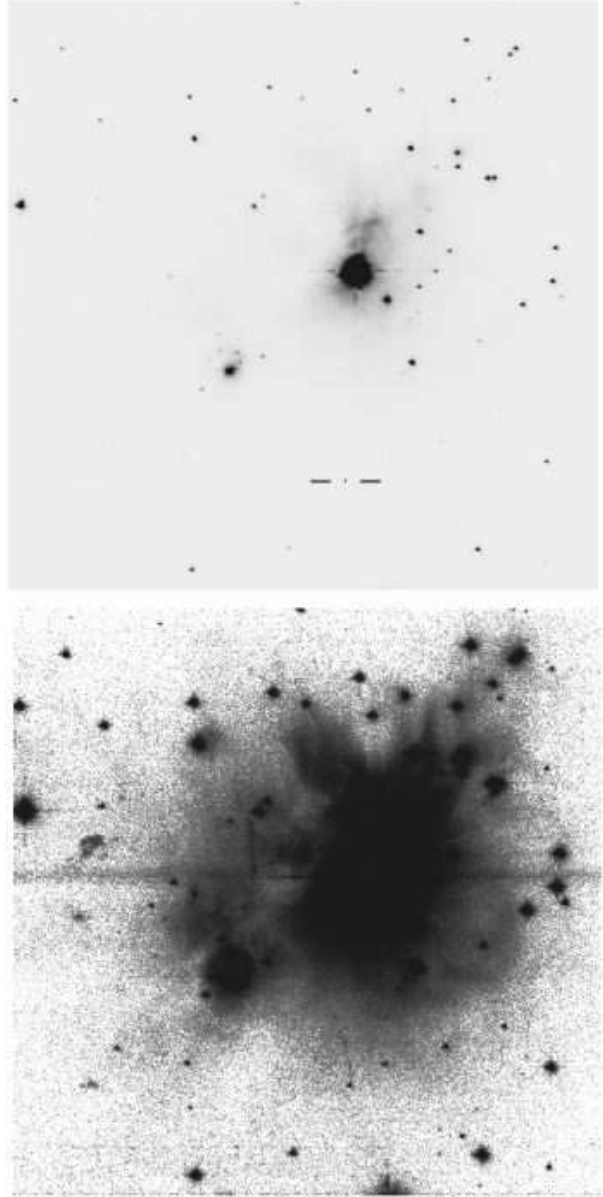}
\caption{Wide-field image obtained with the UH 2.2 m telescope. 
$Top$: Unstretched image.  
The bright object near the center is Allen's object (NGC 2264 IRS; Allen 1972).  
The location of KH 15D is indicated with horizontal tick marks.  
$Bottom$:  Stretched image showing the filament near KH 15D.  
North is up, and east is to the left.}
\end{figure}

\clearpage

\begin{figure}
%\vspace{-2cm}
\epsscale{1.0}
\plotone{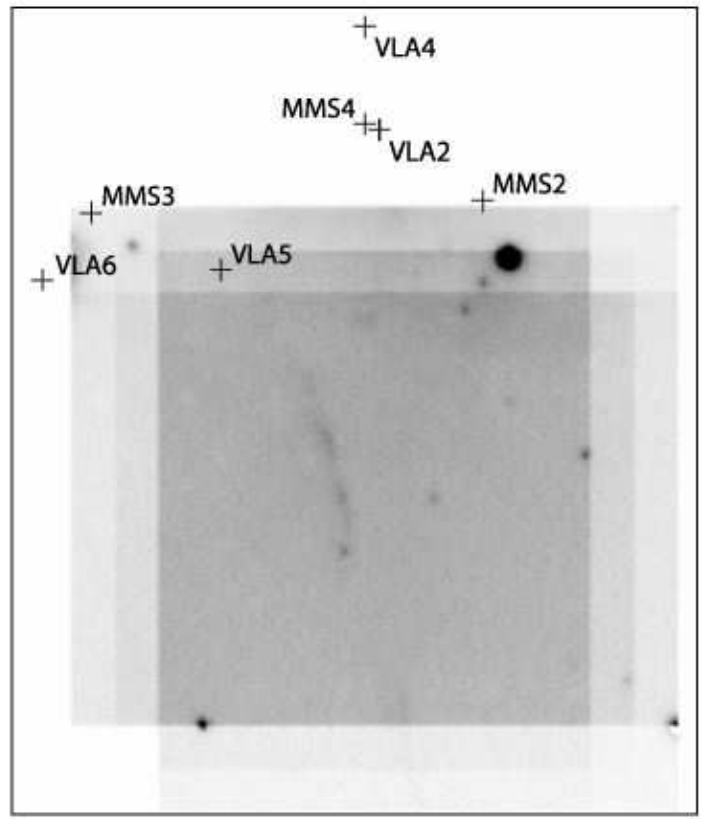}
\caption{Location of VLA and millimeter sources near KH 15D 
(Reipurth et al. 2003; Ward-Thompson et al. 2000) superimposed 
on the image shown in Fig. 1c. 
We show the overlapping images for clarity.  
}
\end{figure}

\clearpage

\begin{deluxetable}{cccc}
%\tabletypesize{\scriptsize}
\tablecaption{Position and Surface Brightness of the Knots \label{tbl-1}}
\tablewidth{0pt}
\tablehead{
\colhead{} & \colhead{}   & \colhead{}   &
\colhead{Surface Brightness}\\
\colhead{Knot} & \colhead{$\Delta$RA}   & \colhead{$\Delta$Dec}   &
\colhead{(mag arcsec$^{-2}$)}
}
\startdata
A & 0\farcs1 E & 6\farcs2 N & 18.6  \\
B & 1\farcs9 E & 13\farcs2 N & 19.0 \\
 \enddata

\tablecomments{
RA and Dec refer to the offsets from KH 15D in arcsec.  
}

\end{deluxetable}


\begin{thebibliography}{}

\bibitem[]{} Agol, E., Barth, A. J., Wolf, S., \& Charbonneau, D. 2003, 
preprint(astro-ph/0309309)
\bibitem[]{} Allen, D. A. 1972, \apj, 172, L55
\bibitem[]{} Barge, P., \& Viton, M. 2003, \apj, 593, L117
\bibitem[]{} Bary, J. S., Weintraub, D. A., \& Kastner, J. H. 2003, \apj, 586, 1136
\bibitem[]{} Deming, D., Charbonneau, D., \& Harrington, J. 2004, \apj, in press.
\bibitem[]{} Eisl{\"o}ffel, J., Mundt, R., Ray, T. P., \& Rodr\'{\i}guez, L. F. 2000, 
in Protostars and Planets IV, ed. V. Mannings, A. P. Boss, \& S. S. Russell 
(Tucson: Univ. of Arizona Press), 815
\bibitem[]{} Hamilton, C. M., Herbst, W., Mundt, R., Bailer-Jones, C. A. L., \& Johns-Krull, C. M. 2003, \apj, 591, L45
\bibitem[]{} Hamilton, C. M., Herbst, W., Shih, C., \& Ferro, A. J. 2001, \apj, 554, L201
\bibitem[]{} Herbst, W., et al. 2002, \pasp, 114, 1167
\bibitem[]{} Hirth, G.A., Mundt, R., \& Solf, J. 1977, \aaps, 126, 437
\bibitem[]{} Kobayashi, N., et al. 2000, Proc. SPIE, 4008, 1056
\bibitem[]{} Micono, M., Davis, C. J., Ray, T. P., Eisloeffel, J., \& Shetrone, M. D. 
1998, \apj, 494, L227
\bibitem[]{} Park, B.-G., Sung, H., Bessell, M. S., \& Kang, Y. H. 2000, \aj, 120, 894
\bibitem[]{} Reipurth, B., \& Bally, J. 2001, \araa, 39, 403
\bibitem[]{} Reipurth, B., Raga, A. C., \& Heathcote, S. 1996, \aap, 311, 989
\bibitem[]{} Reipurth, B., Rodr\'{\i}guez, L. F., Anglada, G., \& Bally, J. 2003, \apj, submitted
\bibitem[]{} Simons, D. A., \& Tokunaga, A. 2002, \pasp, 114, 169
\bibitem[]{} Takami, H., et al. 2003, Proc. SPIE, 4839, 21
\bibitem[]{} Tokunaga, A. T., et al. 1998, Proc. SPIE, 3354, 512
\bibitem[]{} Tokunaga, A. T., Simons, D. A., \& Vacca, W. D. 2002, \pasp, 114, 180
\bibitem[]{} Ward-Thompson, D., Zylka, R., Mezger, P. G., \& Sievers, A. W. 2000, \aap, 355, 1122
\bibitem[]{} Winn, J. N., Garnavich, P. M., Stanek, K. Z., \& Sasselov, D. D. 
2003, \apj, 593, L121.


\end{thebibliography}
\end{document}